\documentclass{pasj00}

\begin{document}
\SetRunningHead{A.\ Yamauchi et al.}
{Maser Disk of the Seyfert 2 IC 2560}
\Received{2009 May 25}
\Accepted{2012 March 29}

\title{Water-Vapor Maser Disk at the Nucleus of the Seyfert 2 Galaxy\\
       IC 2560 and its Distance}

\author{Aya \textsc{Yamauchi},\altaffilmark{1}
        Naomasa \textsc{Nakai},\altaffilmark{2}
        Yuko \textsc{Ishihara},\altaffilmark{3}
        Philip \textsc{Diamond},\altaffilmark{4}
    and Naoko \textsc{Sato}\altaffilmark{5}}
\altaffiltext{1}{Mizusawa VLBI Observatory, National Astronomical Observatory of Japan, \\2-12 Hoshigaoka, Mizusawa, Oshu, Iwate 023-0861}
\email{a.yamauchi@nao.ac.jp}
\altaffiltext{2}{Institute of Physics, University of Tsukuba, 1-1-1 Ten-nodai, Tsukuba, Ibaraki 305-8571}
\altaffiltext{3}{Tsukuba Expo Center, 2-9, Azuma, Tsukuba, Ibaraki 305-0031}
\altaffiltext{4}{CSIRO Astronomy and Space Science, PO Box 76, Epping NSW 1710, Australia}
\altaffiltext{5}{Institute for Education on Space, Wakayama University, 930 Sakaedani, Wakayama, Wakayama 640-8510}

\KeyWords{galaxies: active ---
          galaxies: individual (IC 2560) ---
          galaxies: nuclei ---
          masers
          }

\maketitle

\begin{abstract}
We present the results of single-dish and VLBI observations for the water-vapor masers at the nucleus of the Seyfert 2, IC 2560. 
We monitored velocities of the maser features with the 45-m telescope of the Nobeyama Radio Observatory. 
Using the data of 1995--2006, the velocity drift rate was detected to be $\bar{a} = +2.57 \pm 0.04$ km s$^{-1}$ yr$^{-1}$ on the average for 6 systemic features. 
The Very Long Baseline Array (VLBA) with the Very Large Array (VLA) firstly detected a red-shifted and a blue-shifted maser features of IC 2560, in addition to systemic maser features and a continuum component. 
We propose a maser disk in the nuclear region. 
The systemic and red-shifted features are emitted from a nearly edge-on disk with the position angle of ${\it PA} = -46^{\circ}$, which is almost perpendicular to the galactic disk. 
Assuming the Keplerian rotation, the radii of the maser disk are $r = 0.087$--0.335 pc, and the thickness is $2H \leq 0.025$ pc.
The binding mass is $3.5 \times 10^6 \MO$ at a distance of $D = 26$ Mpc, and the mean volume density within the inner radius is $1.3 \times 10^9 \MO$ pc$^{-3}$, strongly suggesting a massive black hole at the center. 
A continuum component was detected at the 0.2 pc southwest of the disk center, and considered as a jet ejected from the nucleus, with an angle of $70^{\circ}$ from the disk. The blue-shifted maser feature is located on the continuum component, being interpreted to be a ``jet maser". 
The distance to IC 2560 is estimated to be $D = 31^{+12}_{-14}$ Mpc from the geometry of the maser disk and the velocity drift rate.
\end{abstract}

\section{Introduction}
Water-vapor (H$_2$O) maser emission ($J_\mathrm{K_{a}K_{c}} = 6_{16}$--$5_{23}$ rotational transition at 22.23508 GHz) is a unique probe to directly investigate the structure and dynamics of active galactic nuclei (AGN) on the (sub-)parsec scale. 
Since the brightness temperature of H$_2$O maser emission is extremely high ($T_{\rm B} \sim 10^{10}$ K), it is one of a few emission lines that can be observed with very long baseline interferometry (VLBI). 
By using VLBI, the distribution and motion of dense gas in AGN can be measured on scales of (sub-)milliarcseconds (mas), where 1 mas corresponds to 0.05 pc at a distance of 10 Mpc.

Presently, more than 80 galaxies, many of which are type-2 Seyfert or LINER systems, have been known to radiate H$_2$O maser emission from their nuclei. 
In the active galaxy NGC 4258, VLBI observations of the maser emission have revealed the existence of a compact disk in Keplerian rotation and of a massive black hole at its nucleus (\cite{miyo95}). 
Another evidence for a compact edge-on disk rapidly rotating was obtained from a secular velocity drift of the systemic maser features. 
The systemic features of NGC 4258 showed an increase of velocities with about 9 km s$^{-1}$ yr$^{-1}$ by monitoring observations over periods of several years (e.g., \cite{has94}, \cite{green95}, \cite{nakai95}, \cite{herr99}). 
In addition, measurements of the distribution and velocity drifts of the masers led to a direct determination of the distance of the galaxy (\cite{miyo95}; \cite{nakai96}; \cite{herr99}; \cite{argon07}). 
Applying this method to more distant galaxies would make an evaluation of the cosmological constants possible.

Other galaxies also show evidences of circumnuclear disks. 
Position measurements of maser spots and position-velocity diagram analysis by VLBI observations have found rotating edge-on disks at the nuclei of NGC 1068 (\cite{GG97}), NGC 3079 (\cite{T98}; \cite{S00}; \cite{yama04}; \cite{kon05}), NGC 4945 (\cite{green97}), Circinus Galaxy (\cite{green03a}), NGC 3393 (\cite{kon08}), UGC  3789 (\cite{reid09}), IC 1481 (\cite{mamyo09}), and other six active galaxies (\cite{kuo11}).
In addition, some galaxies show secular velocity drifts of the systemic features; NGC 2639 (\cite{wil95}), IC 2560 (\cite{ishi01}), Mrk 1419 (\cite{hen02}), and UGC 3789 (\cite{braatz10}).

Water-vapor masers in IC 2560 have been analyzed by \citet{ishi01}, and here we will re-analyze them more accurately, using new VLBI observations.
IC 2560 is an SB(r)b galaxy (\cite{RC3}) located in the southern sky with a declination of $-33^{\circ}$. 
The galaxy is in the Antila cluster at a distance of 26 Mpc (\cite{aaron89}) and receding with a velocity of $V_{\rm sys} = 2876 \pm 20$ km s$^{-1}$ (\cite{strau92}), which has been converted into the local standard of rest (LSR) frame (from the heliocentric frame; $V_{\rm LSR} = V_{\rm hel} - 12.1$ km s$^{-1}$) and also into the radio definition (velocities will be in the same definition throughout this paper). 
The adopted parameters for IC 2560 in this paper are listed in table \ref{i2para}.
A bar structure with a full length of $120''$ (or 15 kpc), two clear spiral arms, and a box-shaped bulge can be seen in the optical image (see section 4.5). 
The galaxy was classified as a Seyfert 2 from optical line observations (\cite{fair86}). 
The nucleus has a 2--10 keV luminosity of $\sim 1.0 \times 10^{41}$ erg s$^{-1}$ (\cite{ishi01}; \cite{iwa02}; \cite{mad06}; \cite{tilak08}). 
\citet{braatz96} detected H$_2$O maser emission from IC 2560, during their survey for H$_2$O masers in nearby active galaxies. 
The detected emission was near to the systemic velocity of the galaxy, and its peak flux density was 0.19 Jy. 
\citet{ishi01} detected blue- and red-shifted high velocity features, offset from the systemic velocity by $\Delta V = 213$--418 km s$^{-1}$. 
They have also measured velocities of the systemic features in 1996--2000 and detected a secular velocity drift of $2.62 \pm 0.09$ km s$^{-1}$ yr$^{-1}$. 
For the red-shifted feature, no significant velocity drift was able to be seen with the upper limit of 0.5 km s$^{-1}$ yr$^{-1}$ (1$\sigma$) in 1999--2000. 
The blue-shifted features were too weak to measure velocity drift. 
They observed this source with VLBA in 1996 and 1998, and detected the systemic maser features and a continuum component, but did not detect the high velocity features. 
Assuming a compact Keplerian edge-on disk for the maser features, the radius was $r = 0.068$--0.26 pc. 
The binding mass and the mass density within 0.068 pc were estimated to be $2.8 \times 10^6 \MO$ and $2.1 \times 10^9 \MO$ pc$^{-3}$, respectively. 

In this paper, we present results of new VLBA observations concerning all the maser features, and discuss the nuclear structure of IC 2560 based on the results.

\section{Observations}
\subsection{The NRO 45-m Telescope}
We observed H$_2$O maser emission from 1995 June through 2006 April using the 45-m telescope of the Nobeyama Radio Observatory\footnotemark (NRO). 
The half-power beam width and the aperture efficiency of the telescope were $\mathrm{HPBW} = \timeform{73''} \pm \timeform{1''}$ and $\eta_\mathrm{a} = 0.63 \pm 0.02$ at 22 GHz, respectively.
\footnotetext{Nobeyama Radio Observatory is a branch of the National Astronomical Observatory of Japan, National Institutes of Natural Sciences.}

The front-end receivers utilized HEMT amplifiers cooled to 20 K, equipped with two polarized feeds that received right and left-circular polarization simultaneously. 
The system noise temperature, $T_\mathrm{sys}$, including the atmospheric effect and the antenna ohmic loss, was 110--640 K, depending on the weather conditions and the observing elevations of $\timeform{12D}$--$\timeform{21D}$. 
The receiver back-ends were 2048-channel high-resolution Acousto-Optical Spectrometers (AOS). 
Each AOS provided a frequency resolution of 37 kHz and a bandwidth of 40 MHz, which corresponded to a velocity resolution of 0.50 km s$^{-1}$ and a velocity width of 540 km s$^{-1}$ at 22 GHz, respectively. 
We used eight AOS; each one overlapped its neighbors by 5 MHz, resulting in a total velocity coverage of $V_\mathrm{LSR} = 1015$--4430 km s$^{-1}$.

The observations were made in the position-switching mode with an off-source position of $\timeform{5'}$ in the azimuth direction. 
The pointing accuracy was about $\timeform{10''}$. 
The calibration of the line intensity was performed by chopping the sky and a reference load at room temperature, yielding an antenna temperature, $T^{*}_\mathrm{A}$, corrected for the atmospheric attenuation. 
The measured antenna temperature, $T^{*}_\mathrm{A}$, was converted into the flux density, $S$, using the sensitivity, $S/T^{*}_\mathrm{A} = 2.76 \pm 0.09$ Jy K$^{-1}$, calculated from the aperture efficiency.

\begin{figure}[tbp]
 \begin{center}
  \FigureFile(160mm,160mm){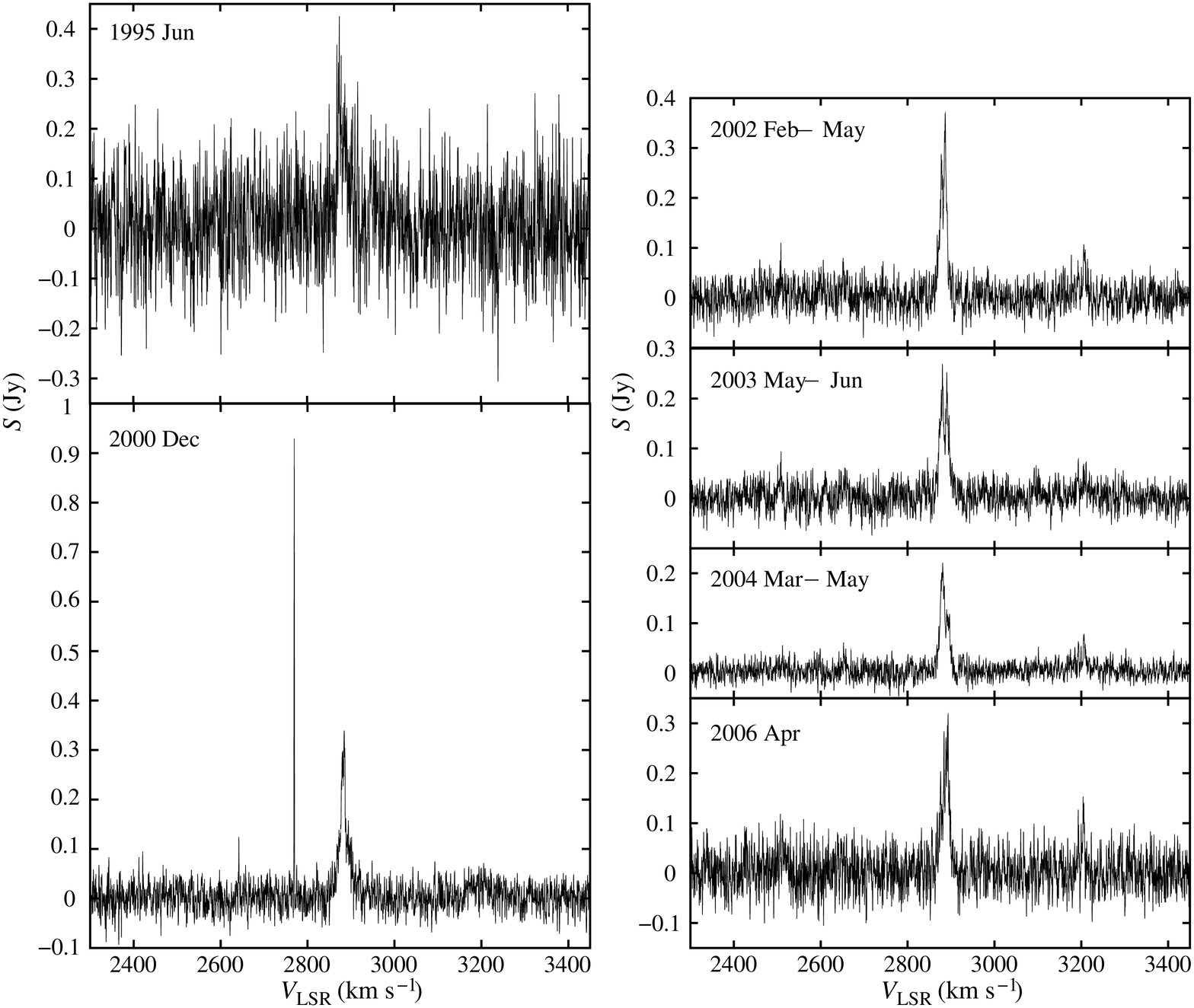}
 \end{center}
 \caption{Maser spectra of IC 2560 measured with the Nobeyama 45-m telescope 
          from 1995 June to 2006 April 
          (as for the spectra from 1996 January to 2000 June, see \cite{ishi01}).
          Each is an averaged spectrum during the period described on the top left. 
          The velocities are with respect to the local standard of rest (LSR) and 
          in the radio definition.
          Note that in panel of 2000 December, a blue-shifted feature 
          at $V_{\rm radio, LSR} = 2769$ km s$^{-1}$ flared up 
          in its flux density.}
 \label{each}
\end{figure}

Figure \ref{each} shows the maser spectra of IC 2560 measured with the 45-m telescope from 1995 June to 2006 April, but those from 1996 January to 2000 June, which have been shown in  \citet{ishi01}, are omitted.
Figure \ref{sp} shows a spectrum of the maser averaged over all observing epochs from 1995 June to 2006 April.

\subsection{VLBA}
The observations of this research were made using  the Very Long Baseline Array (VLBA) (except an antenna at Hancock) and the phased Very Large Array (VLA) of the National Radio Astronomy Observatory\footnotemark (NRAO), USA. 
\footnotetext{The National Radio Astronomy Observatory is a facility of the National Science Foundation operated under cooperative agreement by Associated Universities, Inc.}

IC 2560 was observed on 2000 April 10 for systemic and high velocity maser features. 
Four IFs were recorded, each with a bandwidth of 8 MHz divided into 128 channels (0.84 km s$^{-1}$ velocity resolution). 
The LSR velocities at the band centers of the IFs were 3185, 2880, 2635, and 2490 km s$^{-1}$ (see figure \ref{sp}).

The data were processed on the VLBA correlator at NRAO, and after the correlation, data reduction including calibration and imaging were processed using the Astronomical Image Processing System (AIPS) package. 
The bandpass response was calibrated by observing 4C39.25, and the residual delays and fringe rates were estimated using 1037-295. 
An amplitude calibration was performed on the basis of measured system temperature.
Self-calibration was applied using the strongest maser feature at $V_{\rm LSR} = 2876$ km s$^{-1}$ as a reference. 
The imaging was performed with the CLEAN-method. 
For all maser features, CLEAN maps were made using an intermediate weighting between natural and uniform weighting.
The maser features were imaged by averaging 4 channels (3.36 km s$^{-1}$) for IF 2, or by averaging 6 channels (5.04 km s$^{-1}$) for the other IFs. 

The synthesized beams and the image noise are given in table \ref{beam}.

\section{Results}
\subsection{Velocity Drifts}
\begin{figure}
 \begin{center}
  \FigureFile(130mm,130mm){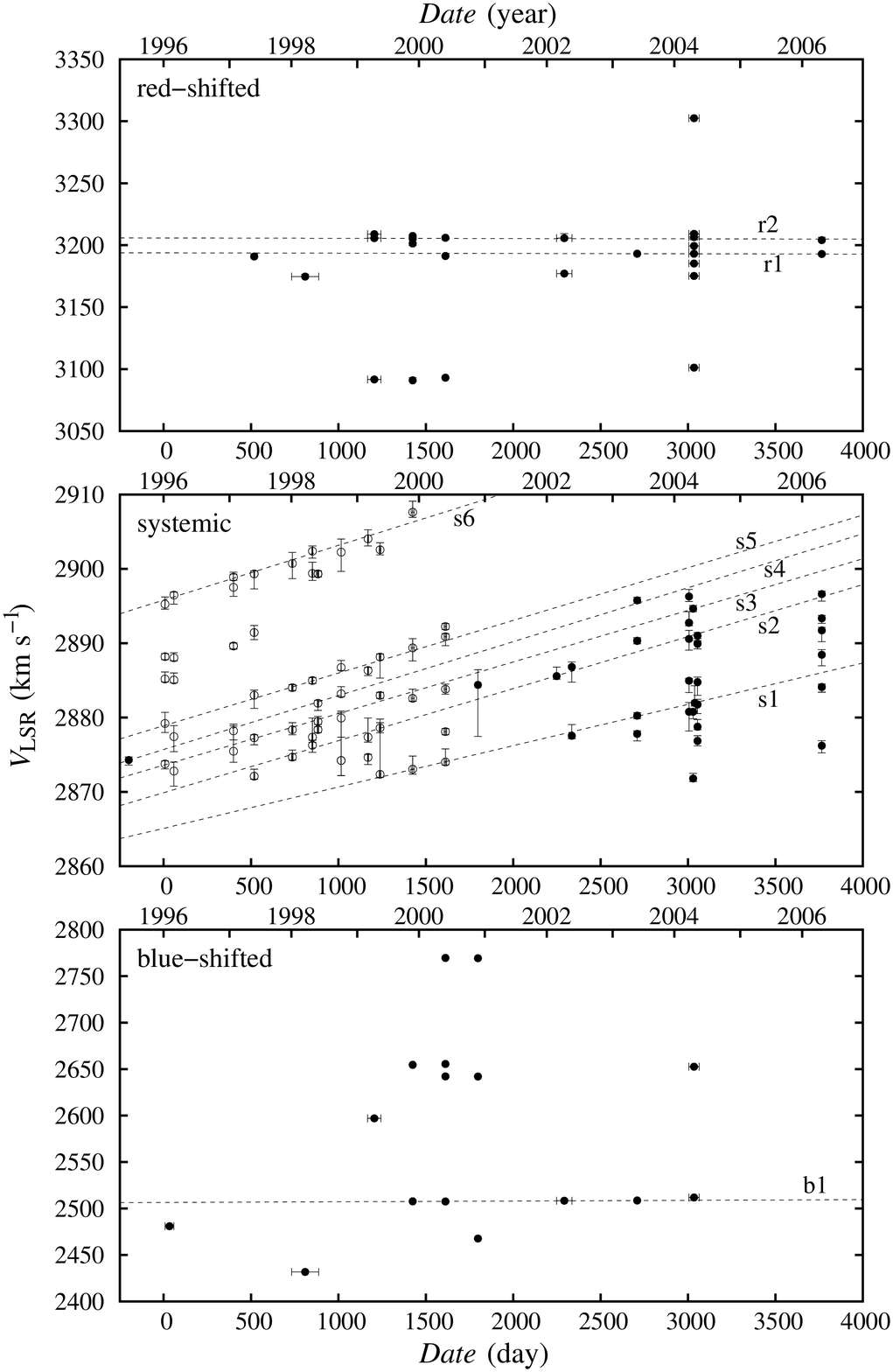}
 \end{center}
 \caption{Velocity variations of the maser features from 1996 January 1. 
          Open circles indicate the data published in \citet{ishi01} and 
          filled circles are our new data.
          The vertical error bars show the velocity range having an intensity  
          of $2 \sigma$ down from each maser peak. 
          The horizontal error bars show the observing epochs of averaged data.
          The dotted lines are the results of least-squares fitting
          for the measured velocities of the maser features, 
          and labels r1, r2, ... , b1 are their numbers 
          (see tables \ref{vel} and \ref{rate}).}
 \label{drift}
\end{figure}

Figure \ref{drift} shows the velocity variations of the red-shifted, systemic, and blue-shifted features at $V_{\rm LSR} = 3050$--3350 km s$^{-1}$, 2860--2910 km s$^{-1}$, and 2400--2800 km s$^{-1}$, respectively, measured with the Nobeyama 45-m telescope. 
All velocity peaks are listed in table \ref{vel}, including the data used in \citet{ishi01}.
For the systemic features, \citet{ishi01} reported those velocities from 1996 January to 2000 May (the open circles in figure \ref{drift}); we newly plot the features with peak intensities of $\geq 5 \sigma$ for each observation in 1995 June and 2000 December to 2006 April. 
The vertical error bars show the range having an intensity of $2 \sigma$ down from each maser peak. 
The best-fit drift rates for 6 systemic features are listed in table \ref{rate}. 
The rates of the velocity drifts are $a = 2.03$--2.69 km s$^{-1}$ yr$^{-1}$, and the weighted average is $\bar{a} = 2.57 \pm 0.04\ \mathrm{km\ s}^{-1}\ \mathrm{yr}^{-1}$, where the error is one sigma. The averaged rate is consistent with and more accurate than the result of \citet{ishi01}.

For the red- and blue-shifted features, we plot the features with peak intensities of $\geq 4 \sigma$. 
We fit 2 red-shifted and 1 blue-shifted feature (see table \ref{rate}), and get negative rates and a positive rate, respectively. 
The spiral shock model (\cite{maoz98}) predicted that red-shifted features have negative rates and blue-shifted features have positive rates. 
Actually, \citet{yama05} detected such rates in NGC 4258. 
Our results for IC 2560 may be consistent with the theoretical prediction, but the measured rates are less than $3 \sigma$ of those fitting errors and we need additional monitoring observations to confirm them. 

\subsection{Distribution of the Maser Features}
Table \ref{position} lists the measured positions of maser spots, and figure \ref{m+c} shows the distribution of the maser spots in the coordinate system relative to the position of the strongest maser feature at $V_{\rm LSR} = 2876$ km s$^{-1}$. 
The circles indicate the positions of the maser peaks with peak intensities of $\geq 5 \sigma$, where $1 \sigma \simeq 11$ or 6 mJy beam$^{-1}$ for the maps 
with averaging 4 or 6 channels, respectively. 
The position errors were $\Delta \theta ({\rm rms}) = 0.5 \theta_{\rm beam} / {\it SNR} = 0.030$--0.099 mas and 0.010--0.036 mas in the major and minor axis directions of the synthesized beam, respectively, where $\theta_{\rm beam}$ is the synthesized beam size and {\it SNR} the signal-to-noise ratio of the peak emission. 

Systemic features, which have a velocity range of 2867--2890 km s$^{-1}$, were located around $(\Delta {\it RA}, \Delta {\it Decl}) \approx (0\ {\rm mas}, 0\ {\rm mas})$, within 0.4 mas (see also figure \ref{radec}). 
The peak flux density was 0.145 Jy beam$^{-1}$. 
The isotropic luminosity of the systemic features was $\sim 45 \LO$ at 26 Mpc. 
A red-shifted feature with the center velocity of $V_{\rm LSR} = 3201$ km s$^{-1}$ was detected at $(\Delta {\it RA}, \Delta {\it Decl}) = (-0.837\ {\rm mas}, 0.774\ {\rm mas})$. 
A spatial separation from the strongest systemic feature was $1.140 \pm 0.030$ mas, corresponding to $0.144 \pm 0.004$ pc. 
The peak flux density was $0.062 \pm 0.007$ Jy beam$^{-1}$, and the isotropic luminosity was $\sim 1 \LO$ at 26 Mpc. 
A blue-shifted feature with the center velocity of $V_{\rm LSR} = 2656$ km s$^{-1}$ was detected at $(\Delta {\it RA}, \Delta {\it Decl}) = (-1.573\ {\rm mas}, 0.131\ {\rm mas})$. 
A spatial separation from the strongest systemic feature was $1.578 \pm 0.047$ mas, corresponding to $0.199 \pm 0.006$ pc. 
The peak flux density was $0.030 \pm 0.006$ Jy beam$^{-1}$, and the isotropic luminosity was $\sim 0.5 \LO$ at 26 Mpc.

Figure \ref{radec} shows an enlarged map of the systemic features in figure \ref{m+c}, with the systemic features detected in 1998 January 2 by \citet{ishi01}.
Assuming that the position of the feature with $V_{\rm LSR} = 2876$ km s$^{-1}$ (without channel average) observed in 1998 was the same as observed in 2000, we plot the maser features; filled and open circles indicate the peak positions of the maser spots detected at the $\geq 5 \sigma$ level in 2000 and in 1998, respectively.
The position errors of the spots detected in 1998 were $\Delta \theta ({\rm rms}) =  0.012$--0.076 mas and 0.004--0.024 mas in the major and minor axis directions of the synthesized beam, respectively. 
The results of a least-squares fitting of the systemic features observed in 1998 and 2000 are ${\it PA} = -8 \pm \timeform{4D}$ and ${\it PA} = \timeform{-19D}^{+12}_{-10} $, respectively, but affected by the synthesized beam elongated toward the north-south direction (table \ref{beam}).

\subsection{Distribution of Continuum Emission}
A 22 GHz continuum component was detected in a map using the data of observations in 1998  (as for the observations, see \cite{ishi01}). 
Its position referred to that of the maser feature at $V_{\rm LSR} = 2876$ km s$^{-1}$. 
We averaged over the channels at a velocity range of 2646--3024 km s$^{-1}$, avoiding maser features. 
To improve the signal-to-noise ratio, a Gaussian taper of 100 M$\lambda$ was applied for the $v$ direction. 
A CLEAN map was made using NATURAL weighting. 
The peak flux density and the integrated flux density were $1.8 \pm 0.3$ mJy beam$^{-1}$ and $1.7 \pm 0.5$ mJy, respectively, and the continuum emission did not extend more than the synthesized beam ($2.17 \times 1.18$ mas). 
Assuming that the peak position of the strongest maser features ($V_{\rm LSR} = 2876$ km s$^{-1}$) observed in 1998 was the same as it observed in 2000, it is possible to overlay the continuum map on the maser map, such as figure \ref{m+c}. 
The peak position of the continuum component was at $(\Delta {\it RA}, \Delta {\it Decl}) = (-1.400\ {\rm mas}, -0.722\ {\rm mas})$. 
A spatial separation from the strongest systemic feature was $1.575 \pm 0.095$ mas, corresponding to $0.199 \pm 0.012$ pc. 
The blue-shifted maser feature detected was located within the 3$\sigma$ contour of the continuum component.

\section{Discussion}
\subsection{Rotating Edge-on Disk}
\begin{figure}
 \begin{center}
  \FigureFile(160mm,160mm){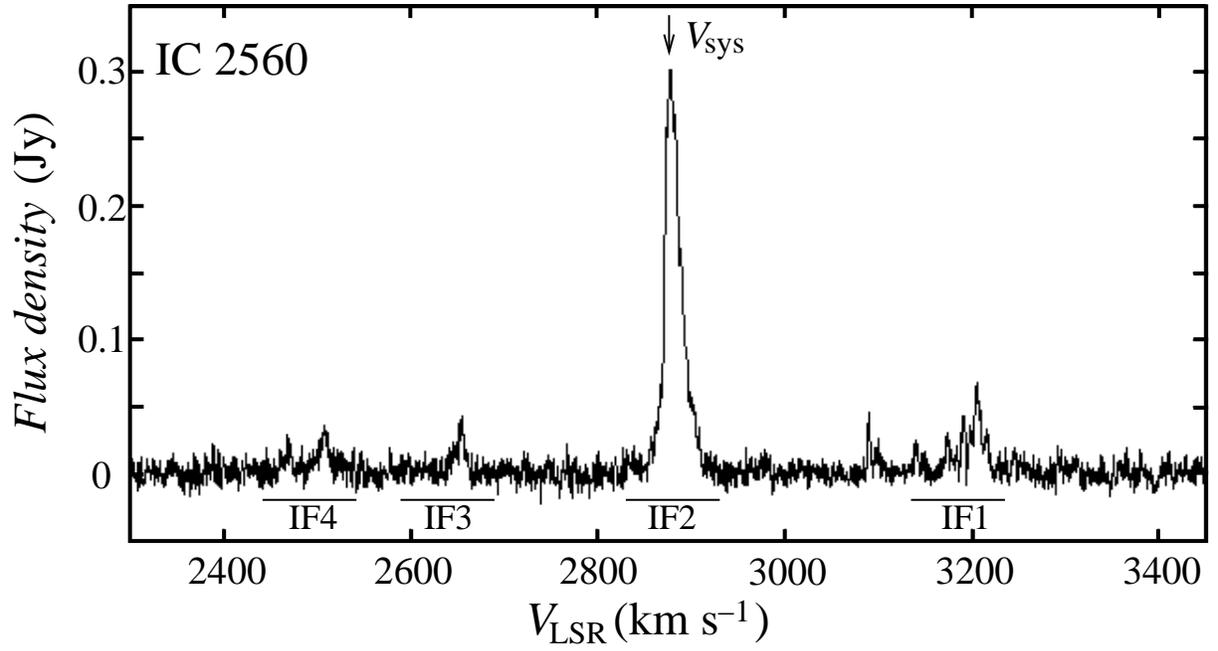}
 \end{center}
 \caption{The water-vapor maser spectrum of IC 2560
          averaged over all observing epochs from 1995 June to 2006 April. 
          The arrow denotes the systemic velocity of the galaxy, 
          $V_{\rm radio, LSR} = 2876$ km s$^{-1}$ (\cite{strau92}). 
          Four thin lines under the spectrum indicate the velocity coverage of 
          VLBA observations in 2000.}
 \label{sp}
\end{figure}

\begin{figure}[tbp]
 \begin{center}
  \FigureFile(160mm,160mm){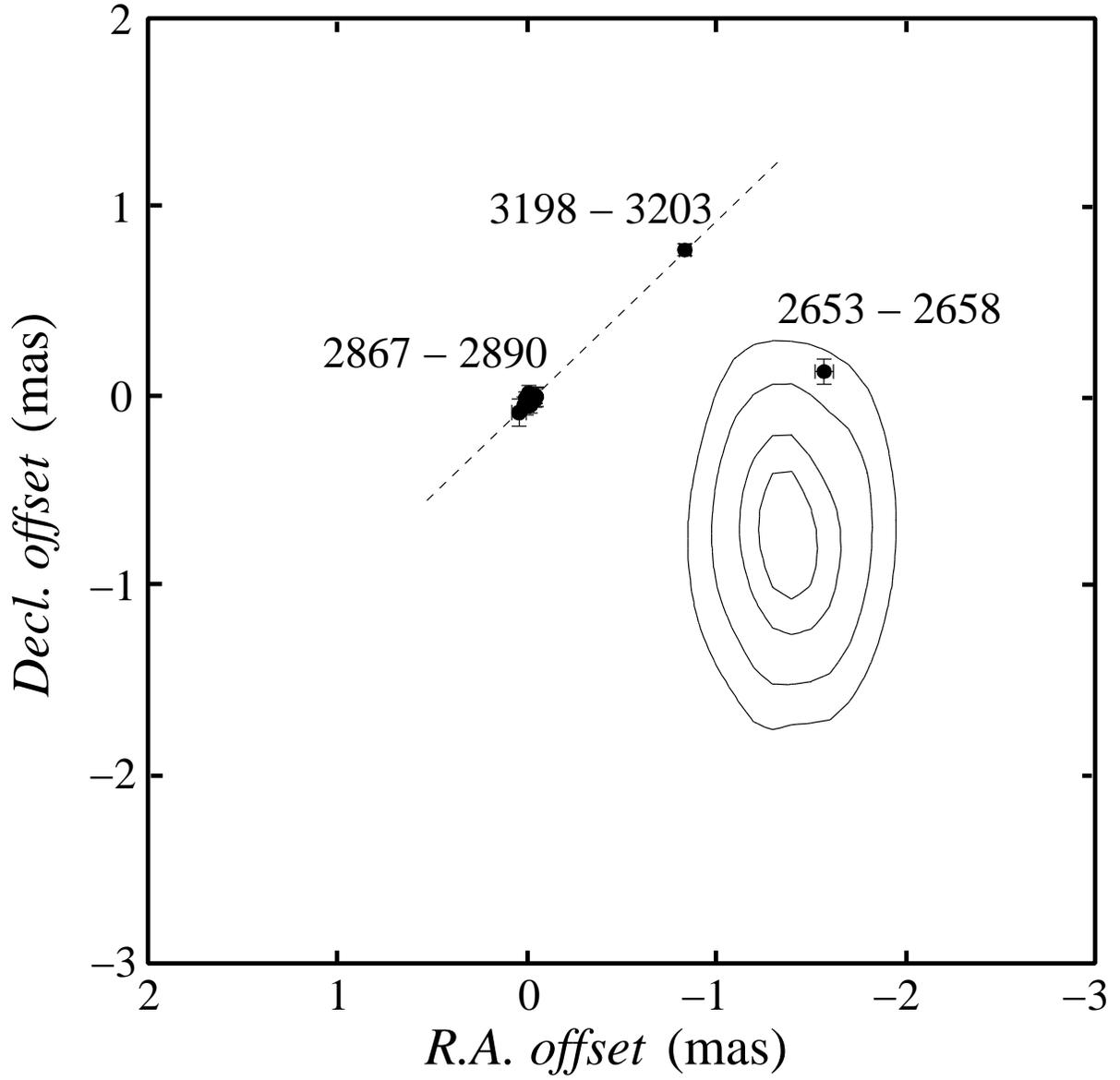}
 \end{center}
 \caption{Distribution of the water masers (observed in 2000) and 
          continuum (1998) emission in the nuclear region of IC 2560. 
          The origin of the coordinates, $(\Delta {\it RA}, \Delta {\it Decl}) 
          = (0\ {\rm mas}, 0\ {\rm mas})$, represents the peak position of 
          the strongest maser feature at $V_{\rm LSR} = 2876$ km s$^{-1}$. 
          Filled circles indicate the peak positions of the maser spots 
          detected at the $\geq 5 \sigma$ level.
          Numbers are the LSR velocity of the maser spots in km s$^{-1}$. 
          Contours indicate a 22 GHz continuum component. 
          The contour levels are 3, 4, 5, and 5.5$\sigma$ 
          ($1 \sigma = 0.31$ mJy beam$^{-1}$). 
          A dashed line with the position angle of ${\it PA} = \timeform{-46D}$ 
          is the result of a least-squares fitting of the circles.
          One mas corresponds to 0.13 pc at the distance of $D = 26$ Mpc.}
 \label{m+c}
\end{figure}

\begin{figure}[tbp]
 \begin{center}
  \FigureFile(80mm,80mm){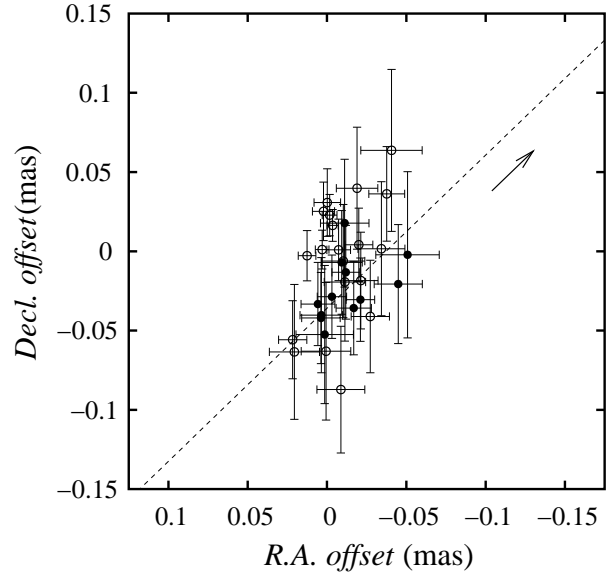}
 \end{center}
 \caption{Distribution of the systemic maser features. 
          The origin of the coordinates is the same as in figure \ref{m+c}. 
          Filled and open circles indicate the peak positions of the maser spots 
          detected at the $\geq 5 \sigma$ level in 2000 and in 1998, respectively. 
          A dashed line with the position angle of ${\it PA}= \timeform{-46D}$ 
          is the same as in figure \ref{m+c}.
          An arrow shows the $+$ direction of the horizontal offset axis 
          in figure \ref{pv}.}
 \label{radec}
\end{figure}

The H$_2$O megamasers in IC 2560 show characteristic natures seen in other megamasers which have an edge-on maser disk rotating around the galactic center; 
(1) the maser spectrum of figure \ref{sp} is composed of the systemic, red- and blue-shifted features, 
(2) most of maser spots except a maser at $(\Delta {\it RA}, \Delta {\it Decl}) = (-1.573\ {\rm mas}, 0.131\ {\rm mas})$ align on the sky (figure \ref{m+c}), 
(3) the velocities of the systemic features show secular drift but high-velocity features do not (figure \ref{drift}), 
and (4) there is a jet-like continuum component nearly perpendicular to the aligned maser distribution. 
From these results, we propose an edge-on disk with the position angle of ${\it PA} = -46 \pm \timeform{1D}$ which is the result of a least-squares fitting of the systemic and red-shifted features (figure \ref{m+c}), rotating around $(\Delta {\it RA}, \Delta {\it Decl}) \approx (0\ {\rm mas}, 0\ {\rm mas})$ which is the position of the systemic velocity of the galaxy. 
The inclination angle of the disk is assumed to be $i \approx \timeform{90D}$.
[The position angle of ${\it PA} = \timeform{-46D}$ is different from the position angles obtained from the systemic features only in section 3.2.
The discrepancy is mainly because that ${\it PA}$ of the systemic features is affected by the synthesized beam elongated toward the north-south direction (table \ref{beam}).]
A red-shifted feature rotates at the radius of $r = 0.144 \pm 0.021$ pc ($1.14 \pm 0.17$ mas) with the velocity of $V_{\rm rot} = |V_{\rm LSR} - V_{\rm sys}| = |3201 - 2876| = 325$ km s$^{-1}$. 
The distribution of the systemic features perpendicular to ${\it PA} = \timeform{-46D}$ gives the thickness of the disk of $2H \leq 0.025$ pc. 

[If the maser disk in IC 2560 has warps, such as NGC 4258 (e.g., \cite{miyo95}) and Circinus Galaxy (\cite{green03a}), its position angle may be different. 
A least-squares fitting of only the systemic features in 1998 and 2000 (see figure \ref{radec}) gives the position angles of ${\it PA} = \timeform{-8 \pm 4D}$ and ${\it PA} = \timeform{-19D}^{+12}_{-10}$, respectively, which are much smaller than ${\it PA} = \timeform{-46D}$, but may be affected by the synthesized beam elongated toward the north-east direction (see table \ref{beam}).
It is difficult, however, to insist the warped disk, because the position angles of the inner and outer disks are indeterminate.]

The blue-shifted feature at $(\Delta {\it RA}, \Delta {\it Decl}) = (-1.573\ {\rm mas}, 0.131\ {\rm mas})$, located within the 3$\sigma$ contour of the 22 GHz continuum component, may not belong to the maser disk but may emit due to stimulation by continuum radiation from the background source (i.e., a jet, see section 4.3). 
NGC 1068 has maser features associated with continuum components (jets), which are called ``the jet masers", apart from ``the disk masers" (\cite{galli01}). 
The Circinus Galaxy showed some maser features associated with a wide-angle outflow from the nucleus with different velocities from the disk masers (\cite{green03a}), and IC 1481 also showed a jet maser as well as disk masers (\cite{mamyo09}). 

The blue-shifted features in the maser disk were not detected with VLBA because of their weakness (figure \ref{sp}), and the red-shifted features around 3090 km s$^{-1}$ were not observed with VLBA. 
To confirm our edge-on disk model, we are conducting further sensitive VLBI observations of these features. 

Continuum emission at the position of the systemic features in figure \ref{m+c} (i.e., the galactic nucleus) was not detected ($< 0.93$ mJy beam$^{-1} = 3 \sigma$).
This is probably because that the continuum emission is attenuated by thermal absorption in a layer of ionized gas above the disk.

\subsection{Super-Massive Black Hole}
The mass inside the radius of the rotating disk, $r = 0.144 \pm 0.021$ pc, 
is given by
\begin{eqnarray}
M &=& \frac{V_{\rm rot}^2 r}{G} = (3.5 \pm 0.5) \times 10^6 \MO,
\end{eqnarray}
where $V_{\rm rot}$ ($= 325$ km s$^{-1}$) is the rotation velocity at $r$, $G$ the gravitational constant, and a spherical distribution of the central matter is assumed.
This value revises a binding mass of $M = 2.8 \times 10^6 \MO$ estimated by \citet{ishi01} using only the velocities in the maser spectrum detected with the Nobeyama 45-m telescope and the drift rates of the velocities of the systemic features. 
This mass is smaller than those of NGC 4258 ($3.9 \times 10^7 \MO$, \cite{miyo95}; \cite{herr99}) by an order of magnitude, but is the same order as those of the Circinus Galaxy ($1.7 \times 10^6 \MO$, \cite{green03a}), NGC 4945 ($1.4 \times 10^6 \MO$, \cite{green97}), and NGC 3079 [(2--3)$\times 10^6 \MO$, \cite{yama04}] .
The mean volume density inside the radius of $r = 0.144 \pm 0.021$ pc is
\begin{eqnarray}
\rho &=& \frac{3 M}{4 \pi r^3} = (2.8 \pm 1.3) \times 10^8 \MO\ {\rm pc}^{-3}.
\end{eqnarray}

Blue- and red-shifted features in figure \ref{sp} were detected at $V_{\rm LSR} = 2458$--2661 km s$^{-1}$ and 3089--3222 km s$^{-1}$, respectively. 
If the high-velocity features except the blue-shifted features around $V_{\rm LSR} = 2656$ km s$^{-1}$ (jet masers) are circularly rotating, the rotation velocities of the edge-on disk are $V_{\rm rot} = |V_{\rm LSR} - V_{\rm sys}| = 215$--418 km s$^{-1}$ for the blue-shifted ($V_{\rm LSR} = 2458$--2556 km s$^{-1}$) and $V_{\rm rot} = 213$--346 km s$^{-1}$ for the red-shifted ($V_{\rm LSR} = 3089$--3222 km s$^{-1}$). 
If the rotation is Keplerian ($V_{\rm rot} \propto r^{-1/2}$) like NGC 4258 (\cite{miyo95}), we can estimate the radii of the maser disk to be $r = (325/V_{\rm rot})^2 \times 0.144 = 0.087$--0.329 pc (0.69--2.60 mas) for the blue-shifted and $r = 0.127$--0.335 pc (1.01--2.65 mas) for the red-shifted, using the radius (0.144 pc) and velocity (325 km s$^{-1}$) of the red-shifted feature at $(\Delta {\it RA}, \Delta {\it Decl}) = (-0.837\ {\rm mas}, 0.774\ {\rm mas})$. 
The blue-shifted features of $V_{\rm LSR} = 2458$ km s$^{-1}$ and the red-shifted features of $V_{\rm LSR} = 3089$ km s$^{-1}$ have the minimum and maximum radii, respectively.
Figure \ref{disk} shows the proposed maser disk. 
The mean volume density inside the radius of $r_{\rm in} = 0.087$ pc is
\begin{eqnarray}
\rho &=& (1.3 \pm 0.6) \times 10^9 \MO\ {\rm pc}^{-3}, 
\end{eqnarray}
which is comparable to that of NGC 4258 [$3.4 \times 10^9 \MO$ pc$^{-3}$, \cite{miyo95}; $4.9 \times 10^{12} \MO$ pc$^{-3}$, \cite{maoz95}; both rescaled to the new distance of 7.2 Mpc determined by \citet{herr99}].
The extremely high density of IC 2560 strongly suggests that this central mass is a super-massive black hole like NGC 4258 [see detailed discussion in \citet{ishi01}]. 

For IC 2560, hard X-ray emission from the nucleus has been observed. 
The 2--10 keV luminosity estimated by correcting for absorption was $L = 1.0^{+1.0}_{-0.4} \times 10^{41}$ erg s$^{-1}$ (\cite{ishi01}; ASCA), $L \lesssim 3 \times 10^{42}$ erg s$^{-1}$ (\cite{iwa02}; {\it Chandra}), $L \gtrsim 3 \times 10^{41}$ erg s$^{-1}$ (\cite{mad06}; {\it Chandra}), and $L = 6.3 \times 10^{41}$ erg s$^{-1}$ (\cite{tilak08}; {\it XMM-Newton}). 
To see how luminous the galaxy is, we compare these X-ray luminosities with the Eddington luminosity. 
The Eddington luminosity $L_{\rm E}$ is a maximum luminosity for a spherically  symmetric object with a given mass. 
If the object radiates luminosity larger than $L_{\rm E}$, then the force from radiation pressure wins over the gravity, and thus gas cannot accrete toward the central object (black hole) but is blown away from the surface. 
The Eddington luminosity for the central mass of $M = (3.5 \pm 0.5) \times 10^{6} M_{\odot}$ in IC 2560 is
\begin{eqnarray}
L_{\rm E} &=& 4 \pi \frac{GMm_{\rm H}c}{\sigma_{\rm T}} = (4.4 \pm 0.7) \times 10^{44}\ {\rm erg\ s^{-1}},
\end{eqnarray}
where $c$ is the velocity of light, $m_{\rm H}$ the mass of a hydrogen atom and $\sigma_{\rm T}$ the Thomson cross section ($\sigma_{\rm T} = 6.65 \times 10^{-25}$ cm$^{2}$). 
So the normalized luminosity becomes $L / L_{\rm E} \sim 10^{-4}$--$10^{-3}$ for the above luminosities, indicating that IC 2560 is actually a low luminosity AGN. 

\subsection{Jet}
A continuum component is located southwest of the rotational center of the maser disk, inclined by $\sim \timeform{20D}$ from the rotation axis. 
Thus we consider the continuum component a jet ejected from the nucleus. 
Jets nearly perpendicular to maser disks have been detected also at the nuclei of NGC 4258 (\cite{herr97}, 1998), NGC 1068 (\cite{galli96}), NGC 3079 (\cite{yama04}), and IC 1481 (\cite{mamyo09}).

A counter jet at the northeastern side of the maser disk was not detected with the upper limit of 0.93 mJy beam$^{-1}$ ($3 \sigma$) which is half of the southwestern jet. 
Absence of the counter jet may be due to attenuation by thermal absorption in a layer of ionized gas above the maser disk, if the disk is slightly inclined. 
Absence of central continuum emission may be also caused by attenuation by the same thermal absorption. 
In this case, the maser disk would incline like figure \ref{disk}, i.e., the northeastern side is nearer to an observer. 
Such relation between continuum emission and a maser disk was seen in NGC 4258 (\cite{herr97}, 1998). 

For IC2560, there has been almost no other VLBI continuum observation at not only 22 GHz but also other frequencies.
More sensitive observations are needed to investigate the jets and expected continuum emission at the nucleus.

\subsection{Distance to IC 2560}
\begin{figure}[tbp]
 \begin{center}
  \FigureFile(160mm,160mm){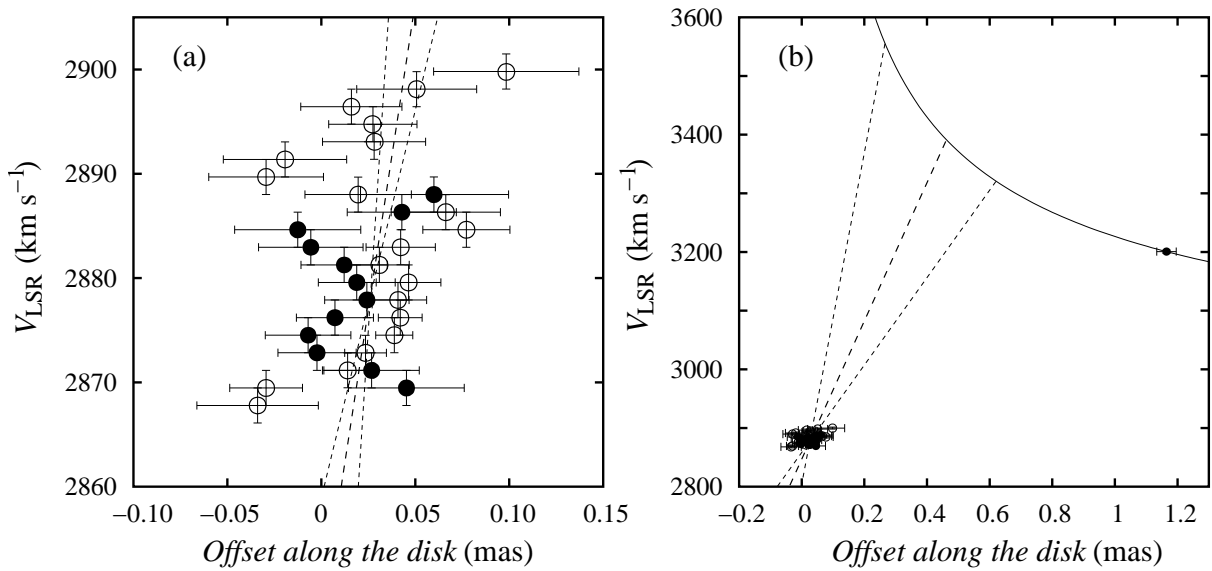}
 \end{center}
 \caption{Position-velocity diagram of (a) the systemic maser features, 
          and (b) the systemic and red-shifted features 
          along the dotted line with ${\it PA} = \timeform{-46D}$ 
          in figures \ref{m+c} and \ref{radec}.
          The abscissas are the offset from the position of the unaveraged 
          maser feature with the systemic velocity of IC 2560, 
          $V_{\rm LSR} = 2876$ km s$^{-1}$.
          The ordinates are the LSR velocities. 
          Filled circles indicate the peak positions and velocities of 
          the systemic and red-shifted features detected in 2000. 
          Open circles indicate the peak positions and velocities of 
          the systemic features detected in 1998. 
          Thick dashed lines are the result of a least-squares fitting of 
          the circles.
          Thin dashed lines show one sigma error of the fitting result.
          A solid curve indicates an assumed Keplerian curve 
          through a red-shifted feature.}
 \label{pv}
\end{figure}

We can estimate the distance to IC 2560, assuming a circular and Keplerian rotation of the maser disk like NGC 4258.
Figure \ref{pv} shows position-velocity diagrams along the maser disk with ${\it PA} = \timeform{-46D}$.
Filled circles indicate the peak positions and velocities of the systemic and the red-shifted features detected in 2000. 
Open circles indicate the peak positions and velocities of the systemic features detected in 1998. 
Thick dashed lines are the result of a least-squares fitting of all the circles (i.e., both of 1998 and 2000).
Thin dashed lines show one sigma error of the fitting result.
A solid curve in (b) indicates an assumed Keplerian curve through a red-shifted feature at $(r, V_{\rm LSR}) = ($1.165 mas, 3201 km s$^{-1}$). 
Intersections of the dashed lines and the Keplerian curve give 
the apparent distance (radius) of the systemic features from the galactic center, $r_{app} = 0.46^{+0.17}_{-0.20}$ mas. 
In the cases of a least-squares fitting of only the open (1998 data) and filled circles (2000 data), $r_{app} = 0.46^{+0.18}_{-0.22}$ mas and $r_{app} = 0.33^{+0.47}_{-0.33}$ mas, respectively.
The three $r_{app}$ are consistent with each other within the errors.
For the radii of all the circles ($r_{app} = 0.46^{+0.17}_{-0.20}$ mas), their rotation velocities are $V_{\rm rot(in)} = |V_{\rm LSR} - V_{\rm sys}| = 515^{+138}_{-59}$ km s$^{-1}$. 

The absolute distance of the systemic features can be estimated using the velocity drift rate $dV_{||}/dt = 2.57 \pm 0.04$ km s$^{-1}$ yr$^{-1}$ (see section 3.1) as the following (e.g., \cite{ishi01}), 
\begin{eqnarray}
r &=& V_{\rm rot(in)}^2 \sin i \left| \frac{dV_{||}}{dt} \right|^{-1} \nonumber \\
  &=& 0.070 \pm 0.003\ {\rm pc},
\end{eqnarray}
where $i$ is the inclination angle of the rotating disk and assumed to be $i \approx \timeform{90D}$, i.e., an edge-on disk. 
Equation (5) is valid under a circular rotation of the maser disk. 
By comparing the apparent (measured using all the circles) and absolute radii, the distance to the host galaxy can be determined to be $D = r/r_{\rm app} = 31^{+12}_{-14}$ Mpc which is consistent with the distance of 26 Mpc estimated using the infrared Tully-Fisher relation (\cite{aaron89}) within the errors, although the errors are large. 
To improve accuracy of the distance and to check the Keplerian rotation, we need VLBI observations with higher sensitivity to detect more masers of both systemic and high velocity features.

\subsection{Independent Rotation of the Nuclear Maser Disk from the Galactic Disk}
\begin{figure}[tbp]
 \begin{center}
  \FigureFile(160mm,160mm){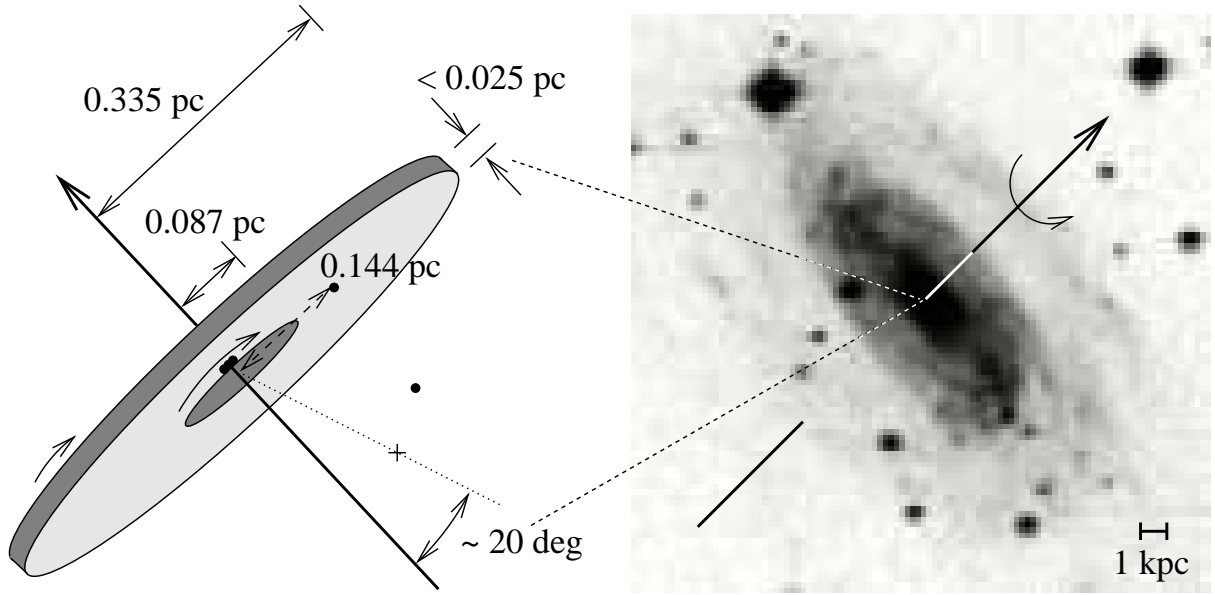}
 \end{center}
 \caption{{\it Left:} Model for the maser disk of IC 2560.
           The rotation velocities at the inner and outer radii are
           418 and 213 km s$^{-1}$, respectively.
           Circles and a cross shows the peak positions of maser spots 
           and a continuum component, respectively. 
           A thick arrow denotes the rotating axis of the maser disk. 
           {\it Right:} Optical image of the galactic disk of IC 2560 
           adopted from Digitized Sky Surveys\footnotemark. 
           \ The scale is $3' \times 3'$ in R.A. and Decl; 
           north is top and east is left.
           The inclination and position angle are $\timeform{63D}$ and 
           $\timeform{45D}$, respectively. 
           Assuming a trailing spiral, the galactic disk rotates as shown 
           in a thin arrow.}
 \label{disk}
\end{figure}

The right panel of figure \ref{disk} is an optical image of IC 2560.
The inclination and position angles of the large-scale galactic disk 
have been measured to be $i = \timeform{63D}$ and ${\it PA} = \timeform{45D}$, respectively, 
from an {\it I}-band image (\cite{mathew92}).
A velocity curve of the kilo-pc scale along ${\it PA} = \timeform{90D}$ showed that radial velocities in the west side were larger than those in the east side (\cite{schu03}), and thus the large-scale disk rotates as the north east side approaches and the south west side recedes.
Assuming that galactic spiral arms are trailing, the large-scale disk rotates as shown with a thin arrow in the right panel.
On the other hand, the rotation axis of the compact maser disk (${\it PA} = \timeform{-46D}$) is almost perpendicular to that of the galactic disk as shown the left panel of figure \ref{disk}.

Much difference of the rotating axis of the nuclear maser disk and the galactic disk of IC 2560 is reminiscent of the case of NGC 4258, whose difference of the position angle between the rotation axes of the galactic and maser disks is $\sim \timeform{110D}$ (\cite{miyo95}), indicating reverse rotation to each other. 
Also UGC 3789 (\cite{reid09}) has the large difference of the position angle between the rotation axes of the galactic and maser disk.
Detailed statistical analysis of the relation between the galactic and maser disk rotation among AGN megamasers will be made in the forthcoming paper.

\section{Summary}

H$_2$O maser emission from the nucleus of IC 2560 was observed with a single-dish and VLBI. 
The results are summarized as follows: 
\begin{enumerate}
\item The velocity drift rates were measured to be $\bar{a} = +2.57 \pm 
      0.04$ km s$^{-1}$ yr$^{-1}$ and $-0.09 \pm 0.15$ km s$^{-1}$ yr$^{-1}$ 
      as for six systemic features and two red-shifted features, respectively, 
      and $a = +0.28 \pm 0.23$ km s$^{-1}$ yr$^{-1}$ for a blue-shifted feature. 
\item Not only the systemic features but also the red- and blue-shifted 
      features were detected with VLBI.
      The velocities for the red- and blue-shifted features are 
      3201 km s$^{-1}$ and 2656 km s$^{-1}$, respectively.
      The velocities relative to the galactic systemic velocity were 
      325 km s$^{-1}$ for a red-shifted feature, and 220 km s$^{-1}$ 
      for a blue-shifted feature. 
      The isotropic luminosities of the blue-shifted, red-shifted, and 
      systemic features were $L = 0.5 L_{\odot}$, $1L_{\odot}$, 
      and $45 L_{\odot}$, respectively, at the distance of 26 Mpc.
\item The systemic and red-shifted features were emitted from a nearly edge-on 
      disk with the position angle of ${\it PA} \approx \timeform{-46D}$. 
      The thickness of the maser disk is $2H \leq 0.025$ pc. 
      The binding mass is $(3.5 \pm 0.5) \times 10^6 \MO$. 
      Assuming the Keplerian rotation and combining with the single-dish 
      spectrum, the radii become to be $r = 0.087$--0.335 pc. 
      The mean volume density within the inner radius is $(1.3 \pm 0.6) \times 
      10^9 \MO$ pc$^{-3}$, strongly suggesting existence of a super-massive 
      black hole.
\item The rotating axis of the maser disk of IC 2560 was nearly perpendicular 
      to that of the galactic disk.
\item A 22 GHz continuum component was detected at the southwest of the disk 
      center by $0.199 \pm 0.012$ pc, 
      and considered as a jet ejected from the nucleus. 
      A position angle of the component relative to the maser disk was 
      ${\it PA} \sim \timeform{70D}$. 
      The blue-shifted maser feature was located on the continuum component, 
      and thus is interpreted to be a ``jet maser".
\item The distance to IC 2560 is estimated to be $D = 31^{+12}_{-14}$ Mpc
      from geometry of the maser disk and the velocity drift rate, 
      assuming the circular and Keplerian rotation of the disk.
\end{enumerate}

In our observations with VLBA, only limited high velocity features in the maser disk were detected. We are conducting more sensitive VLBI observations of other high velocity features, both red- and blue-shifted features, as well as the systemic features to obtain the rotation curve without the assumption of a Keplerian rotation, and thus to exactly determine the distance of IC2560.

\bigskip

We thank NISHIYAMA Kohta, and HIROTA Akihiko for helping with the maser observations using the Nobeyama 45-m telescope. 
We also thank members of NRO for their continuous support.


\clearpage
\onecolumn

\clearpage

\footnotetext
{The Digitized Sky Surveys were produced at the Space Telescope Science
 Institute under U.S. Government grant NAG W-2166. The images of
 these surveys are based on photographic data obtained using
 the Oschin Schmidt Telescope on Palomar Mountain 
 and the UK Schmidt Telescope.}

\clearpage

\begin{table}
\begin{center}
\caption{Adopted parameters of IC 2560}
\label{i2para}
\begin{tabular}{ll}
\hline
\hline
Center position\footnotemark[$*$]	   & $\alpha_{\rm J2000.0} = \timeform{10h16m18s.710}$ \\
					   & $\delta_{\rm J2000.0} = \timeform{-33D33'49''.74}$ \\
Morphological type\footnotemark[$\dagger$] & SB(r)b \\
Distance\footnotemark[$\ddagger$]	   & $26$ Mpc \\
Systemic velocity\footnotemark[$\S \#$]	   & $2876 \pm 20$ km s$^{-1}$ \\
Inclination angle\footnotemark[$\|$]	   & $\timeform{63D}$  \\
Position angle\footnotemark[$\|$]	   & $\timeform{45D}$ \\
\hline
\end{tabular}
\end{center}
\footnotemark[$*$] \citet{ishi01}. \\
\footnotemark[$\dagger$] \citet{RC3}. \\
\footnotemark[$\ddagger$] \citet{aaron89}. \\
\footnotemark[$\S$] \citet{strau92}. \\
\footnotemark[$\|$] \citet{mathew92}. \\
\footnotemark[$\#$] Radio definition, and with respect to the local standard of rest. \\
\end{table}

\begin{table}
\begin{center}
\caption{Beam sizes and rms noise of VLBI images for IC 2560.}
\label{beam}
\begin{tabular}{lcccccc}
\hline
\hline
		& Year	& Channels$^{\ast}$	& Major	& Minor	& P.A.	& Rms noise \\
		&		&			& (mas)	& (mas)	& (deg)	& (mJy beam$^{-1}$) \\
\hline
Spectral line	& 2000	& 4	& 0.81	& 0.27	& $-7$	& 10.9 \\
		& 2000	& 6	& 1.00	& 0.36	& $-18$	& \ 6.2 \\
		& 1998\footnotemark[$\dagger$]	& 4	& 1.00	& 0.32	& $-10$	& \ 4.4 \\
Continuum	& 1998\footnotemark[$\dagger$]	&	 	& 2.17	& 1.18	& 4	& \ 0.31 \\
\hline
\end{tabular}
\end{center}
$^{\ast}$ Numbers of averaged channels used when imaging. \\
\footnotemark[$\dagger$] The data of \citet{ishi01} are re-analyzed. 
\end{table}

\begin{longtable}{r@{$\,\pm\,$}llc|r@{$\,\pm\,$}llc}
\caption{Velocities of the maser peaks. }
\label{vel}
\hline
\hline
\multicolumn{2}{c}{Date\footnotemark[$*$]}	& $V_\mathrm{LSR}$\footnotemark[$\dagger$]	& Verr\footnotemark[$\ddagger$]	& \multicolumn{2}{c}{Date\footnotemark[$*$]}	& $V_\mathrm{LSR}$\footnotemark[$\dagger$]	& Verr\footnotemark[$\ddagger$] \\
\multicolumn{2}{c}{(day)}	& (km s$^{-1}$)		& (km s$^{-1}$)	& \multicolumn{2}{c}{(day)}	& (km s$^{-1}$)		& (km s$^{-1}$) \\
\hline
\endhead
\hline
\endfoot
\hline
\multicolumn{8}{l}{\hbox to 0pt{\parbox{180mm}{
\footnotemark[$*$] Days from 1996 January 1 and ranges of the observing epochs of averaged data.\\
 $I$ indicates the data published in \citet{ishi01}. \\
\footnotemark[$\dagger$] LSR velocities of the maser peaks. Superscripts such as b1 show the maser features \\
used to measure drift rates (see figure \ref{drift} and table \ref{rate}).\\
\footnotemark[$\ddagger$] Errors from the velocity range having an intensity of $2 \sigma$ down from each maser peak. \\
}}}

\endlastfoot
\multicolumn{3}{l}{Blue-shifted}	&	      & \multicolumn{3}{l}{}			& \\
1425.0	& 9.0	& 2507.72$^{\rm b1}$	& -0.68/+0.95 & 33.5	& 25.5	& 2481.01		& -0.14/+1.55 \\
1612.0	& 7.0	& 2507.52$^{\rm b1}$	& -1.21/+0.96 & 1612.0	& 7.0	& 2655.54		& -0.95/+0.41 \\
2292.0	& 44.0	& 2508.36$^{\rm b1}$	& -0.41/+0.41 & 1612.0	& 7.0	& 2769.60		& -0.20/+0.08 \\
2709.5	& 4.5	& 2508.61$^{\rm b1}$	& -0.41/+0.41 & 1798.5	& 0.5	& 2467.65		& -0.41/+0.41 \\
33.5	& 25.5	& 2481.01		& -0.14/+1.55 & 1798.5	& 0.5	& 2642.02		& -0.14/+0.68 \\
809.5	& 77.5	& 2431.78		& -0.41/+0.68 & 1798.5	& 0.5	& 2769.34		& -0.14/+0.14 \\
1205.0	& 38.0	& 2597.03		& -0.41/+0.68 & 3034.5	& 29.5	& 2511.87		& -0.41/+0.41 \\
1425.0	& 9.0	& 2654.62		& -0.68/+0.68 & 3034.5	& 29.5	& 2652.53		& -0.41/+0.95 \\
1612.0	& 7.0	& 2642.21		& -0.41/+0.41 & \multicolumn{3}{l}{}			& \\
\hline
\multicolumn{3}{l}{Systemic}		&	      & \multicolumn{3}{l}{}			& \\
1238.5	& 4.5$^I$& 2872.37$^{\rm s1}$	& -0.41/+7.43 & 1612.0	& 7.0$^I$& 2890.88$^{\rm s5}$	& -1.22/+0.41 \\
1425.0	& 9.0$^I$& 2873.06$^{\rm s1}$	& -0.68/+1.76 & 8.0	& 0.0$^I$& 2895.26$^{\rm s6}$	& -0.68/+0.95 \\
1612.0	& 7.0$^I$& 2874.02$^{\rm s1}$	& -0.41/+1.76 & 59.0	& 0.0$^I$& 2896.47$^{\rm s6}$	& -1.22/+0.41 \\
2335.5	& 0.5	& 2877.55$^{\rm s1}$	& -0.41/+1.50 & 400.0	& 0.0$^I$& 2898.88$^{\rm s6}$	& -0.41/+0.41 \\
2709.5	& 4.5	& 2880.26$^{\rm s1}$	& -0.41/+0.41 & 518.5	& 2.5$^I$& 2899.32$^{\rm s6}$	& -2.03/+0.41 \\
3038.0	& 1.0	& 2881.96$^{\rm s1}$	& -0.41/+2.31 & 736.5	& 4.5$^I$& 2900.72$^{\rm s6}$	& -2.03/+1.49 \\
3054.5	& 9.5	& 2881.76$^{\rm s1}$	& -1.22/+0.68 & 851.0	& 2.5$^I$& 2902.39$^{\rm s6}$	& -0.95/+0.68 \\
736.5	& 4.5$^I$& 2874.68$^{\rm s2}$	& -0.41/+0.95 & 1015.5	& 0.5$^I$& 2902.24$^{\rm s6}$	& -2.57/+1.76 \\
851.0	& 2.5$^I$& 2876.27$^{\rm s2}$	& -0.95/+0.41 & 1169.5	& 2.5$^I$& 2904.03$^{\rm s6}$	& -0.95/+1.22 \\
1169.5	& 2.5$^I$& 2877.37$^{\rm s2}$	& -0.68/+2.57 & -200.0	& 3.5	& 2874.29		& -0.68/+0.41 \\
1238.5	& 4.5$^I$& 2878.62$^{\rm s2}$	& -0.68/+0.68 & 8.0	& 0.0$^I$& 2885.19		& -0.41/+0.95 \\
2248.0	& 0.0	& 2885.57$^{\rm s2}$	& -0.41/+1.22 & 8.0	& 0.0$^I$& 2888.19		& -0.41/+0.41 \\
2335.5	& 0.5	& 2886.80$^{\rm s2}$	& -2.04/+0.68 & 59.0	& 0.0$^I$& 2885.05		& -0.41/+0.95 \\
3006.0	& 1.0	& 2890.57$^{\rm s2}$	& -1.50/+1.14 & 59.0	& 0.0$^I$& 2888.04		& -0.41/+0.68 \\
3054.5	& 9.5	& 2891.01$^{\rm s2}$	& -0.41/+0.41 & 400.0	& 0.0$^I$& 2889.63		& -0.41/+0.41 \\
3765.0	& 0.0	& 2896.61$^{\rm s2}$	& -0.95/+0.41 & 400.0	& 0.0$^I$& 2897.52		& -1.22/+2.03 \\
8.0	& 0.0$^I$& 2873.77$^{\rm s3}$	& -0.68/+0.41 & 518.5	& 2.5$^I$& 2872.11		& -0.41/+0.95 \\
59.0	& 0.0$^I$& 2872.80$^{\rm s3}$	& -2.03/+1.22 & 518.5	& 2.5$^I$& 2891.43		& -0.95/+0.95 \\
400.0	& 0.0$^I$& 2875.48$^{\rm s3}$	& -1.49/+3.65 & 851.0	& 2.5$^I$& 2877.36		& -0.41/+2.57 \\
518.5	& 2.5$^I$& 2877.28$^{\rm s3}$	& -0.95/+0.41 & 851.0	& 2.5$^I$& 2899.40		& -0.95/+1.49 \\
736.5	& 4.5$^I$& 2878.36$^{\rm s3}$	& -0.68/+0.95 & 883.0	& 3.5$^I$& 2878.38		& -0.41/+0.41 \\
883.0	& 3.5$^I$& 2879.47$^{\rm s3}$	& -0.41/+0.68 & 883.0	& 3.5$^I$& 2899.33		& -0.41/+0.41 \\
1015.5	& 0.5$^I$& 2879.94$^{\rm s3}$	& -2.57/+0.68 & 1015.5	& 0.5$^I$& 2874.22		& -2.03/+6.62 \\
1238.5	& 4.5$^I$& 2882.98$^{\rm s3}$	& -0.41/+0.41 & 1169.5	& 2.5$^I$& 2874.64		& -0.95/+0.41 \\
1425.0	& 9.0$^I$& 2882.59$^{\rm s3}$	& -0.41/+1.22 & 1238.5	& 4.5$^I$& 2902.57		& -0.68/+0.95 \\
1612.0	& 7.0$^I$& 2883.81$^{\rm s3}$	& -0.68/+0.68 & 1425.0	& 9.0$^I$& 2907.61		& -0.68/+1.49 \\
1798.5	& 0.5	& 2884.40$^{\rm s3}$	& -6.94/+2.04 & 1612.0	& 7.0$^I$& 2878.10		& -0.41/+0.41 \\
3029.5	& 1.5	& 2894.65$^{\rm s3}$	& -0.41/+0.41 & 1612.0	& 7.0$^I$& 2892.24		& -0.41/+0.41 \\
59.0	& 0.0$^I$& 2877.43$^{\rm s4}$	& -2.03/+1.49 & 2709.5	& 4.5	& 2877.81		& -0.95/+0.41 \\
400.0	& 0.0$^I$& 2878.21$^{\rm s4}$	& -1.22/+0.68 & 2709.5	& 4.5	& 2890.32		& -0.41/+0.41 \\
883.0	& 3.5$^I$& 2881.92$^{\rm s4}$	& -0.95/+0.41 & 3006.0	& 1.0	& 2880.78		& -2.59/+1.22 \\
1015.5	& 0.5$^I$& 2883.20$^{\rm s4}$	& -0.41/+0.95 & 3006.0	& 1.0	& 2884.94		& -1.58/+0.33 \\
2709.5	& 4.5	& 2895.77$^{\rm s4}$	& -0.41/+0.41 & 3006.0	& 1.0	& 2892.75		& -0.41/+1.50 \\
3006.0	& 1.0	& 2896.29$^{\rm s4}$	& -0.68/+0.95 & 3029.5	& 1.5	& 2871.80		& -0.41/+0.68 \\
8.0	& 0.0$^I$& 2879.21$^{\rm s5}$	& -1.22/+1.49 & 3029.5	& 1.5	& 2880.78		& -0.95/+0.41 \\
518.5	& 2.5$^I$& 2883.00$^{\rm s5}$	& -1.76/+0.68 & 3054.5	& 9.5	& 2876.87		& -0.68/+0.68 \\
736.5	& 4.5$^I$& 2884.02$^{\rm s5}$	& -0.41/+0.41 & 3054.5	& 9.5	& 2878.77		& -0.41/+0.95 \\
851.0	& 2.5$^I$& 2884.98$^{\rm s5}$	& -0.41/+0.41 & 3054.5	& 9.5	& 2884.76		& -1.77/+0.68 \\
1015.5	& 0.5$^I$& 2886.74$^{\rm s5}$	& -0.41/+0.95 & 3054.5	& 9.5	& 2889.93		& -0.68/+0.41 \\
1169.5	& 2.5$^I$& 2886.34$^{\rm s5}$	& -0.68/+0.41 & 3765.0	& 0.0	& 2876.21		& -0.95/+0.68 \\
1238.5	& 4.5$^I$& 2888.15$^{\rm s5}$	& -2.84/+0.41 & 3765.0	& 0.0	& 2884.10		& -0.68/+0.41 \\
1425.0	& 9.0$^I$& 2889.38$^{\rm s5}$	& -1.76/+1.22 & 3765.0	& 0.0	& 2888.45		& -1.50/+0.68 \\
\multicolumn{3}{l}{Systemic}		&	      & \multicolumn{3}{l}{}			& \\
3765.0	& 0.0	& 2891.72		& -1.50/+0.41 & 3765.0	& 0.0	& 2893.35		& -0.68/+0.41 \\
\hline
\multicolumn{3}{l}{Red-shifted}		&	      & \multicolumn{3}{l}{}			& \\
2709.5	& 4.5	& 3193.15$^{\rm r1}$	& -0.68/+0.41 & 1425.0	& 9.0	& 3090.96		& -0.68/+2.22 \\
3034.5	& 29.5	& 3193.11$^{\rm r1}$	& -0.95/+1.23 & 1425.0	& 9.0	& 3201.41		& -1.77/+0.68 \\
3765.0	& 0.0	& 3192.90$^{\rm r1}$	& -0.41/+0.41 & 1425.0	& 9.0	& 3207.39		& -0.41/+0.68 \\
1205.0	& 38.0	& 3205.64$^{\rm r2}$	& -0.68/+0.68 & 1612.0	& 7.0	& 3093.04		& -0.41/+0.41 \\
1425.0	& 9.0	& 3205.22$^{\rm r2}$	& -0.41/+0.95 & 1612.0	& 7.0	& 3191.26		& -0.41/+0.41 \\
1612.0	& 7.0	& 3205.95$^{\rm r2}$	& -1.50/+0.95 & 2292.0	& 44.0	& 3177.11		& -0.41/+0.41 \\
2292.0	& 44.0	& 3205.68$^{\rm r2}$	& -1.22/+3.67 & 3034.5	& 29.5	& 3101.16		& -0.41/+0.41 \\
3034.5	& 29.5	& 3206.44$^{\rm r2}$	& -2.04/+0.95 & 3034.5	& 29.5	& 3175.16		& -0.41/+0.41 \\
3765.0	& 0.0	& 3204.05$^{\rm r2}$	& -0.68/+2.31 & 3034.5	& 29.5	& 3185.22		& -0.68/+0.68 \\
518.5	& 2.5	& 3190.86		& -0.41/+2.05 & 3034.5	& 29.5	& 3199.37		& -0.68/+0.41 \\
809.5	& 77.5	& 3174.71		& -0.41/+1.77 & 3034.5	& 29.5	& 3209.16		& -0.41/+1.50 \\
1205.0	& 38.0	& 3091.64		& -0.41/+0.14 & 3034.5	& 29.5	& 3302.48		& -0.41/+0.68 \\
1205.0	& 38.0	& 3208.91		& -0.41/+0.41 & \multicolumn{3}{l}{}			& \\
\end{longtable}

\begin{table}
\begin{center}
\caption{Drift rates of the maser features.}
\label{rate}
\begin{tabular}{lccr@{$\,\pm\,$}lc}
\hline
\hline
 & & $V_\mathrm{LSR}$\footnotemark[$*$]	& \multicolumn{2}{c}{$a$}	& Average \\
 & & (km s$^{-1}$) & \multicolumn{2}{c}{(km s$^{-1}$ yr$^{-1}$)}& (km s$^{-1}$ yr$^{-1}$) \\
\hline
Blue-shifted	& b1 & 2506.5	& $+0.28$ & 0.23	& \\
\hline
Systemic	& s1 &  2865.1	& $+2.03$ & 0.24	& $+2.57 \pm 0.04$ \\
		& s2 &  2869.9	& $+2.55$ & 0.08	& \\
		& s3 &  2873.6	& $+2.53$ & 0.07	& \\
		& s4 &  2875.7	& $+2.64$ & 0.07	& \\
		& s5 &  2878.9	& $+2.59$ & 0.27	& \\
		& s6 &  2895.8	& $+2.69$ & 0.32	& \\
\hline
Red-shifted	& r1 &  3193.8	& $-0.09$ & 0.23	& $-0.09 \pm 0.15$\\
		& r2 &  3204.9	& $-0.08$ & 0.20	& \\
\hline
\end{tabular}
\end{center}
\footnotemark[$*$] LSR velocities of the maser peaks on 1996 January 1, which is the result of fitting to the data in 1995--2006. \\
Note: Labels b1, s1, ... r2 correspond to the dotted lines in figure \ref{drift} (see also table \ref{vel}).\\
\end{table}

\begin{table}
\begin{center}
\caption{Positions of maser features.}
\label{position}
\begin{tabular}{cccr@{$\,\pm\,$}lr@{$\,\pm\,$}lr@{$\,\pm\,$}l}
\hline
\hline
Year	& $V_\mathrm{LSR}$\footnotemark[$*$]	& Channels\footnotemark[$\dagger$]	& \multicolumn{2}{c}{Flux density}	& \multicolumn{2}{c}{$\Delta {\it RA}$\footnotemark[$\ddagger$]}	& 	\multicolumn{2}{c}{$\Delta {\it Decl}$\footnotemark[$\ddagger$]}	\\
		& (km s$^{-1}$)	& 	& \multicolumn{2}{c}{(Jy beam$^{-1}$)}	& \multicolumn{2}{c}{(mas)}	& 	\multicolumn{2}{c}{(mas)}	\\
\hline
2000	& 3201	& 6	& 0.062	& 0.007	& -0.837	& 0.029	&  0.774	& 0.032 \\
		& 2888	& 4	& 0.066	& 0.011	& -0.051	& 0.020	& -0.002	& 0.052 \\
		& 2886	& 4	& 0.074	& 0.010	& -0.045	& 0.015	& -0.021	& 0.038 \\
		& 2885	& 4	& 0.088	& 0.012	&  0.001	& 0.018	& -0.052	& 0.044 \\
		& 2883	& 4	& 0.104	& 0.012	&  0.004	& 0.013	& -0.040	& 0.036 \\
		& 2881	& 4	& 0.133	& 0.012	& -0.017	& 0.011	& -0.036	& 0.029 \\
		& 2880	& 4	& 0.133	& 0.010	& -0.021	& 0.009	& -0.031	& 0.026 \\
		& 2878	& 4	& 0.131	& 0.010	& -0.012	& 0.009	& -0.013	& 0.030 \\
		& 2876	& 4	& 0.141	& 0.011	& -0.003	& 0.009	& -0.029	& 0.026 \\
		& 2875	& 4	& 0.128	& 0.012	&  0.004	& 0.012	& -0.042	& 0.029 \\
		& 2873	& 4	& 0.119	& 0.009	&  0.006	& 0.011	& -0.033	& 0.026 \\
		& 2871	& 4	& 0.117	& 0.011	& -0.009	& 0.013	& -0.007	& 0.033 \\
		& 2869	& 4	& 0.083	& 0.010	& -0.011	& 0.015	&  0.018	& 0.040 \\
		& 2656	& 6	& 0.030	& 0.006	& -1.573	& 0.047	&  0.131	& 0.067 \\
\hline
1998	& 2900	& 4	& 0.043	& 0.005	& -0.041	& 0.019	&  0.064	& 0.051 \\
		& 2898	& 4	& 0.047	& 0.004	& -0.034	& 0.015	&  0.002	& 0.042 \\
		& 2896	& 4	& 0.057	& 0.004	& -0.027	& 0.012	& -0.041	& 0.035 \\
		& 2895	& 4	& 0.056	& 0.004	& -0.021	& 0.011	& -0.018	& 0.030 \\
		& 2893	& 4	& 0.043	& 0.004	& -0.010	& 0.014	& -0.006	& 0.036 \\
		& 2891	& 4	& 0.040	& 0.004	&  0.001	& 0.016	& -0.063	& 0.043 \\
		& 2890	& 4	& 0.039	& 0.004	& -0.009	& 0.015	& -0.087	& 0.040 \\
		& 2888	& 4	& 0.041	& 0.004	& -0.011	& 0.013	& -0.019	& 0.037 \\
		& 2886	& 4	& 0.039	& 0.004	& -0.019	& 0.013 &  0.040	& 0.038 \\
		& 2885	& 4	& 0.046	& 0.003	& -0.038	& 0.011	&  0.036	& 0.030 \\
		& 2883	& 4	& 0.069	& 0.004	& -0.020	& 0.009	&  0.004	& 0.023 \\
		& 2881	& 4	& 0.085	& 0.004	& -0.007	& 0.007	&  0.001	& 0.019 \\
		& 2880	& 4	& 0.096	& 0.005	& -0.000	& 0.008	&  0.031	& 0.021 \\
		& 2878	& 4	& 0.132	& 0.006	&  0.002	& 0.007	&  0.025	& 0.018 \\
		& 2876	& 4	& 0.186	& 0.006	& -0.002	& 0.005	&  0.023	& 0.013 \\
		& 2875	& 4	& 0.208	& 0.005	& -0.003	& 0.004	&  0.016	& 0.010 \\
		& 2873	& 4	& 0.211	& 0.006	&  0.003	& 0.004	&  0.001	& 0.012 \\
		& 2871	& 4	& 0.153	& 0.005	&  0.013	& 0.006	& -0.003	& 0.016 \\
		& 2869	& 4	& 0.069	& 0.004	&  0.022	& 0.009	& -0.056	& 0.025 \\
		& 2868	& 4	& 0.039	& 0.004	&  0.021	& 0.016	& -0.063	& 0.043 \\
\hline
\end{tabular}
\end{center}
\footnotemark[$*$] Center velocities of the maser features. \\
\footnotemark[$\dagger$] Numbers of averaged channels used when imaging. Velocity resolution of one channel was 0.84 km s$^{-1}$.\\
\footnotemark[$\ddagger$] Offsets from the position of the strongest maser feature at $V_{\rm LSR} = 2876$ km s$^{-1}$ (one channel imaging). Because we use only one channel with $V_{\rm LSR} = 2876$ km s$^{-1}$ for self-calibration, the origin of the coordinates is the peak position of the strongest maser feature in one channel map.\\
\end{table}
\end{document}